\documentclass{article}
\usepackage[preprint]{spconf}
\usepackage{amsmath,graphicx}
\usepackage{xcolor}
\usepackage{multirow}
\usepackage{amssymb}
\usepackage{longtable}
\usepackage{tabularx}
\usepackage[hidelinks]{hyperref}
\usepackage{enumitem}

\title{CONSTRUCTING COMPOSITE FEATURES FOR INTERPRETABLE MUSIC-TAGGING\thanks{Accepted at the 2026 IEEE International Conference on Acoustics, Speech, and Signal Processing (ICASSP 2026), Barcelona, Spain, May 4--8, 2026.}\thanks{\textcopyright~2026 IEEE. Personal use of this material is permitted. Permission from IEEE must be obtained for all other uses, in any current or future media, including reprinting/republishing this material for advertising or promotional purposes, creating new collective works, for resale or redistribution to servers or lists, or reuse of any copyrighted component of this work in other works.}}
\name{%
\begin{tabular}{@{}c@{}}
Chenhao Xue$^{1}$ \qquad Weitao Hu$^{2}$ \qquad
Joyraj Chakraborty$^{1}$ \qquad Zhijin Guo$^{1}$\\
Kang Li$^{1}$ \qquad Tianyu Shi$^{3}$ \qquad Martin Reed$^{4}$ \qquad
Nikolaos Thomos$^{4}$\thanks{The authors thank Puyu Wang for providing music theory validation of the evolved GP expressions.}
\end{tabular}
}

\address{$^{1}$University of Oxford \quad $^{2}$Independent Researcher \quad
$^{3}$University of Toronto \quad $^{4}$University of Essex}

\begin{document}
\def\ninept{\def\baselinestretch{1.004}\let\normalsize\small\normalsize}
\ninept
\maketitle
\begin{abstract}
\vspace{-0.35\baselineskip}
Combining multiple audio features can improve the performance of music tagging, but common deep learning-based feature fusion methods often lack interpretability. To address this problem, we propose a Genetic Programming (GP) pipeline that automatically evolves composite features by mathematically combining base music features, thereby capturing synergistic interactions while preserving interpretability. This approach provides representational benefits similar to deep feature fusion without sacrificing interpretability. Experiments on the MTG-Jamendo and GTZAN datasets demonstrate consistent improvements compared to state-of-the-art systems across base feature sets at different abstraction levels. It should be noted that most of the performance gains are noticed within the first few hundred GP evaluations, indicating that effective feature combinations can be identified under modest search budgets. The top evolved expressions include linear, nonlinear, and conditional forms, with various low-complexity solutions at top performance aligned with parsimony pressure to prefer simpler expressions. Analyzing these composite features further reveals which interactions and transformations tend to be beneficial for tagging, offering insights that remain opaque in black-box deep models.
\end{abstract}
\vspace{-3mm}\section{Introduction\vspace{-0.5mm}}
\label{sec:intro}

\vspace{-3mm}
Music audio tagging concerns automatically assigning descriptive labels or ``tags'' (e.g., mood, theme, instrument) to music tracks based on their audio content. This is a fundamental problem in music information retrieval (MIR) because accurate tags enable the efficient organization and retrieval of large music collections~\cite{won2021semi}. For decades, automatic tagging has been approached via handcrafted feature extraction followed by traditional machine-learning classifiers~\cite{tzanetakis2002musical, prockup2015modeling, zhao2022violinist, lyberatos2024perceptual}. These engineered features were designed to capture low- and mid-level acoustically and musically significant characteristics (e.g., MFCC for timbre or chroma features for tonality), which made them inherently interpretable as they correspond to perceptual aspects of music. In recent years, there has been a shift towards minimal feature engineering and end-to-end deep learning models that learn features directly from audio~\cite{Puig2018EndtoendLF, pons2019musicnn, kong2020panns}. These methods significantly enhance the accuracy of music tagging models with overwhelming parameters, vast labeled data, and synergy modelling of individual music features (closer to human perception~\cite{harrison2020simultaneous, mcpherson2020perceptual}) with explicit or implicit complex feature fusion~\cite{won2021semi, chang2024iiof, liu2025research}. 

While deep-learning models achieve state-of-the-art tagging performance, their opacity remains a concern, as they cannot easily explain why a certain tag was assigned. This is exacerbated in tasks with subjective or ambiguous ground truth ~\cite{buisson2022ambiguity}. The subjective nature of tags like ``happy'' or ``aggressive'' means that even ground truth annotations can vary between annotators~\cite{buisson2022ambiguity, koops2019annotator}. A black-box model could latch onto spurious correlations. Without interpretability, neither developers nor users can tell if the model is leveraging unwarranted biases (e.g., associating ``sadness'' only with slow tempo)~\cite{mishra2017local}. Indeed, a lack of transparency in the tagging decisions makes it difficult to detect dataset biases or flaws in model reasoning. This is critical in music machine learning (ML) where models often train on limited genres or cultures, and hidden biases may lead to poor generalization or unfair outcomes~\cite{holzapfel2018ethical}. 

These practical considerations call for interpretable music audio tagging methods that could approach some advantages of deep learning models, such as the modelling of synergetic effects of music features, which has received limited study in music tasks~\cite{cui2006exploring, makinen2012evolutionary}. In this paper, we introduce a method to automatically construct interpretable composite features from individual base features for music tagging using Genetic Programming (GP), which is a form of symbolic regression~\cite{mei2022explainable, yang2024design}. Our specific contributions are:

\begin{itemize}[itemsep=0.0mm,topsep=0.2mm]
    \item we propose a Genetic Programming pipeline that constructs interpretable composite features by combining base music features to improve tagging and reveal interaction insights;
    \item we demonstrate that the GP-augmented feature sets consistently improve tagging accuracy across two datasets and different types of base features, with gains achieved under modest evaluation budgets;
    \item we analyze the resulting symbolic expressions to identify effective feature interactions and transformations, offering interpretability and insight not available in black-box models.
\end{itemize}

The GP-constructed composite features, expressions, and code for this paper are made available on GitHub\footnote{\url{https://github.com/ChenHX111/GP-Music-Tagging}}.

\vspace{-4mm}\section{Methods}
\label{sec:methods}

\vspace{-3mm}
\begin{figure}[t]
\centering
\begin{minipage}[b]{\linewidth}
  \centering
  \includegraphics[width=\linewidth]{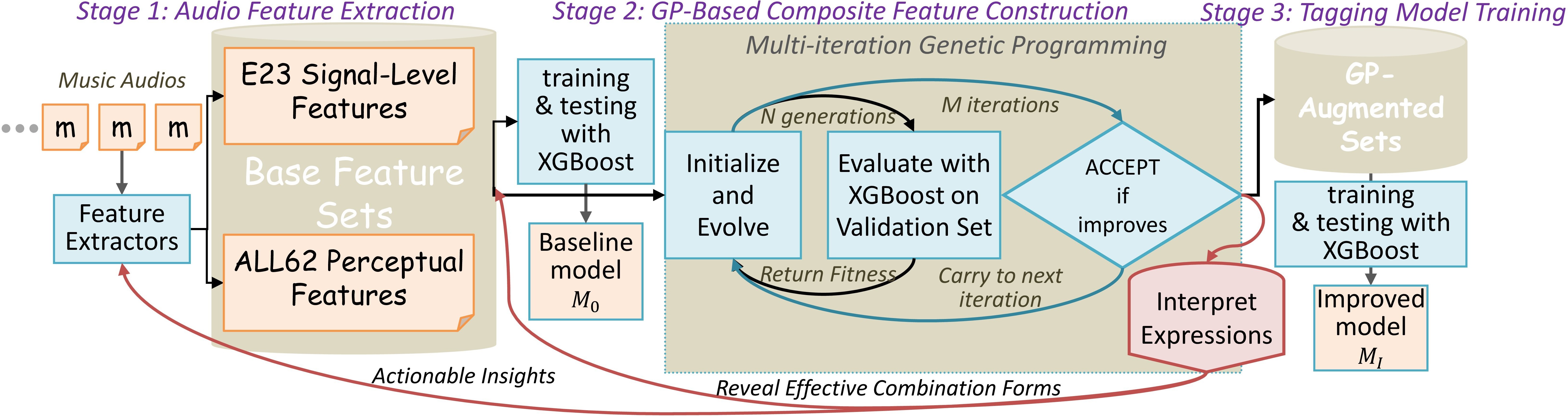}
  \caption{GP pipeline for Composite Feature and Model Construction. \vspace{-6mm}}
  \label{fig:overview}
\end{minipage}
\end{figure}
Although deep learning-based music tagging methods achieve state-of-the-art music-tagging performance, their feature fusion remains opaque, and hence, making it difficult to understand why a tag has been allocated to a piece of music. Traditional handcrafted approaches provide interpretability through feature importance, but cannot systematically discover complex feature combinations at scale. To address this challenge, we propose a GP pipeline that evolves interpretable mathematical combination expressions, providing explicit symbolic representations of feature interactions while enabling automated discovery of effective combinations. As shown in Figure~\ref{fig:overview}, the pipeline consists of three stages: (1) Audio Feature Extraction, (2) GP-Based Composite Feature Construction, and (3) Tagging Model Training.

\vspace{-4mm}\subsection{Audio Feature Extraction\vspace{-2mm}}
We employ two feature sets at different abstraction levels, following established music information retrieval (MIR) practices that combine signal descriptors with perceptual knowledge~\cite{lyberatos2024perceptual}. Specifically,

\textbf{Signal-level Features (E23):} We extract 23-dimensional audio descriptors using the Essentia library's music feature extractor~\cite{bogdanov2013essentia}. This base feature set captures signal-level characteristics (e.g., Loudness, BPM, Onset Rate, Zero Crossing Rate, Dynamic Complexity, Pitch Salience, and Spectral-Centroid);

\textbf{Low and Mid-Level Perceptual Features (ALL62):} We use a 62-dimensional feature set from Lyberatos et al.'s study~\cite{lyberatos2024perceptual}. This set includes: (a) the E23 features described above, (b) 32 ontology-grounded harmonic-function features computed using the Omnizart chord recognition Python library~\cite{wu2021omnizart} mapped to Functional Harmony Ontology classes with normalized $n$-gram frequencies queried via SPARQL~\cite{seaborne2008sparql}, and (c) 7 perceptual features estimated through multi-output regression using a VGG-style Convolutional Neural Networks (CNN) processing Mel-Frequency Cepstral Coefficients (MFCC) segments derived from 15-sec clips. 

Before further processing, all features are scaled to have zero mean and unit variance. These base feature sets are input for GP-based composite feature construction. 

\vspace{-4mm}\subsection{GP-Based Composite Feature Construction\vspace{-2mm}}
\begin{figure}[t]
\centering
\begin{minipage}[b]{\linewidth}
  \centering
  \includegraphics[width=0.9\linewidth]{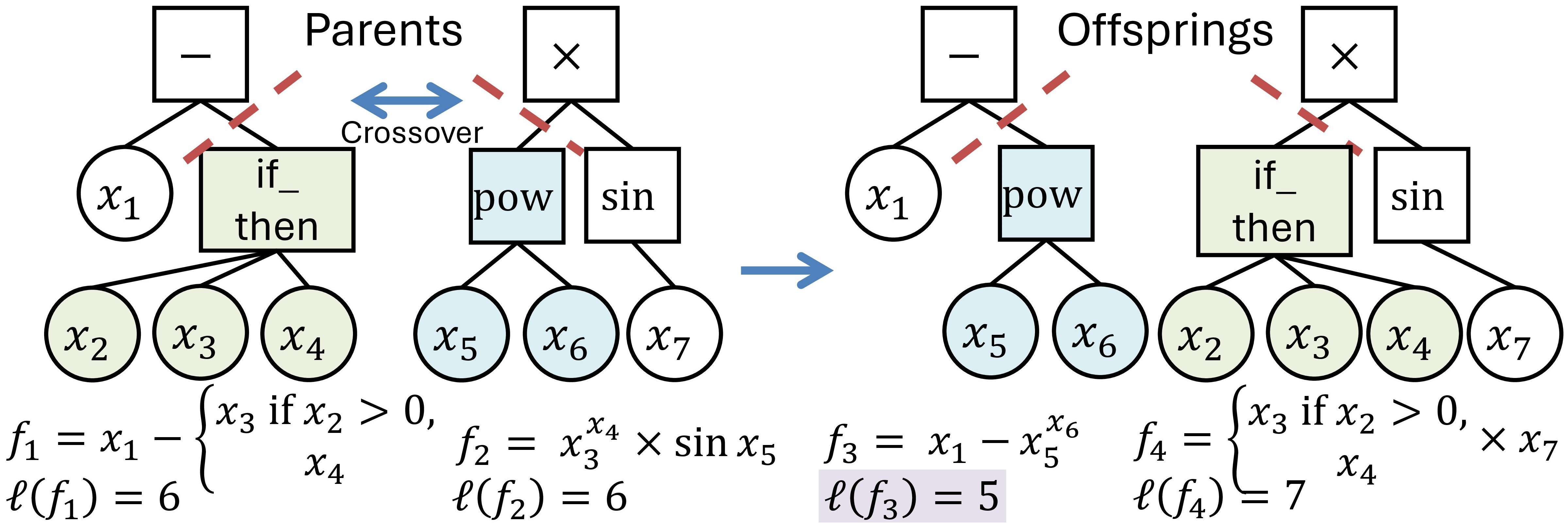}
  \caption{GP crossover operation for composite feature construction\vspace{-4mm}}
  \label{fig:crossover}
\end{minipage}\vspace{-4mm}
\end{figure}

Let us define base features $\mathbf{X} = \{x_1, \ldots, x_n\}$ and target labels $\mathbf{y}$. We first formulate the GP composite feature construction as an optimization problem:\vspace{-3mm}
$$\max_{f_1, \ldots, f_M \in \mathcal{F}} P(\mathbf{X} \cup \{f_1(\mathbf{X}), \ldots, f_M(\mathbf{X})\}, \mathbf{y}) \;-\; \lambda \sum_{i=1}^{M} \ell(f_i)$$\vspace{-3mm} \\ 
where $f_i: \mathbb{R}^n \rightarrow \mathbb{R}$ are composite features from function space $\mathcal{F}$ defined by GP operations $\mathcal{O}$. $P(\cdot, \mathbf{y})$ represents the tagging performance metric. $\ell(f_i)$ is the expression size (number of nodes, i.e., operators and terminals) in the expression tree of $f_i$ (see Figure \ref{fig:crossover}). Our GP system evolves interpretable mathematical expressions that solve this optimization problem at a scale, generating human-readable expressions that reveal feature interactions. Once evolved, these expressions require only basic computations for new instances, concentrating computational cost at training time rather than deployment.

We evolve up to $M$ composite features in $M$ iterations using base features as primitives. Each iteration consists of a full GP run. In our framework, each GP \emph{individual} encodes a scalar composite feature as a rooted expression tree, constructed from standardized base features together with ephemeral constants sampled from $c \sim \mathcal{U}[-2,2]$. The operation set $\mathcal{O}$ includes arithmetic and protected numeric operators (e.g., \texttt{log}, \texttt{sqrt}, \texttt{div}, \texttt{inv}) for linear/nonlinear combination; trigonometric and hyperbolic functions for periodic patterns; \texttt{min}/\texttt{max}; neural-style activations (e.g., \texttt{sigmoid}, \texttt{RELU}, \texttt{LRELU}, \texttt{swish}); and a ternary \texttt{if\_then} for piecewise dependencies, with closure ensuring real-valued outputs. Initialization uses ramped half-and-half (depth in [1,3]). Population evolution uses one-point subtree crossover (with rate 0.8, illustrated in Figure~\ref{fig:crossover}); uniform subtree mutation (rate 0.1) replacing a randomly selected subtree with a full tree of depth in [0,2]; tournament selection (size 3); and a static height limit of 6 to control bloat. Numerical robustness is enforced via protected primitives, fitness penalties for invalid values (\texttt{NaN}/$\infty$), sanitization via \texttt{nan\_to\_num}, and standardization of each candidate feature before evaluation. 

The fitness function evaluates candidates by augmenting the feature set and measuring XGBoost~\cite{chen2016xgboost} held-out validation set performance, thereby directly optimizing for predictive utility. Following parsimony pressure~\cite{poli2008parsimony}, we apply a complexity penalty ($\lambda = 0.01$ per node found by experimentation) to promote interpretability by favoring simpler expressions over complex ones (e.g., $f_3$ over $f_4$ in Figure~\ref{fig:crossover}), preventing bloat and maintaining transparency. The evolved expressions provide interpretability through two mechanisms: first, each expression explicitly shows how base features combine at scale revealing various interactions unavailable in both black-box models~\cite{chang2024iiof, liu2025research} and beyond systematic exploration capabilities of manual feature engineering~\cite{lyberatos2024perceptual, lyberatos2025challenges}, and second, iterative single-feature addition directly assesses each composite feature's contribution.

\vspace{-4mm}\subsection{Tagging Model Training \vspace{-2mm}}

We employ XGBoost~\cite{chen2016xgboost} as our tagging classifier due to its strong performance on tabular data and robustness to mixed feature types. MTG-Jamendo mood/theme tagging is treated as multiple binary classification tasks (multilabel)~\cite{bogdanov2019mtg}, while GTZAN genre tagging uses multiclass classification~\cite{sturm2013gtzan}. We evaluate using ROC-AUC for MTG-Jamendo and accuracy for GTZAN, consistent with established interpretable~\cite{lyberatos2024perceptual,lyberatos2025challenges} and deep learning methods~\cite{kang2025towards, 10097236}.



\vspace{-3mm}\section{Experimental results}
\label{sec:exp_res}

\newcommand{\scoreci}[3]{#1\,{\scriptsize[\,#2--#3\,]}}

\begin{table}[t]
\caption{GP-Augmented vs. Baseline Performance for deployment models}
\label{tab:performance_comparison}
\centering
\renewcommand{\arraystretch}{1}
\setlength{\tabcolsep}{1.75pt} 
\small
\begin{tabular}{l|cc|cc}
\hline
\multirow{2}{*}{\textbf{Method}} & \multicolumn{2}{c|}{\textbf{MTG-Jamendo (AUC)}} & \multicolumn{2}{c}{\textbf{GTZAN (ACC)}} \\
 & \textbf{Score [95\% CI]} & \textbf{$+\Delta$} & \textbf{Score [95\% CI]} & \textbf{$+\Delta$} \\
\hline
ALL62~\cite{lyberatos2024perceptual,lyberatos2025challenges} & 0.727 & -- & 0.765 & -- \\
ALL62 + \textbf{GP100}  & \scoreci{0.729}{0.724}{0.733} & 0.002 & \scoreci{0.800}{0.760}{0.845} & 0.035 \\
ALL62 + \textbf{GP500}  & \scoreci{0.730}{0.724}{0.736} & 0.003 & \scoreci{0.805}{0.760}{0.850} & 0.040 \\
\hline
E23~\cite{lyberatos2024perceptual,lyberatos2025challenges}   & 0.719 & -- & 0.740 & -- \\
E23 + \textbf{GP100}    & \scoreci{0.722}{0.716}{0.728} & 0.003 & \scoreci{0.785}{0.730}{0.830} & 0.045 \\
E23 + \textbf{GP500}    & \scoreci{0.724}{0.717}{0.731} & 0.005 & \scoreci{0.790}{0.735}{0.840} & 0.050 \\
\hline
\end{tabular}
\vspace{-3mm}
\end{table}

\begin{figure}[t]
\centering
\begin{minipage}[b]{0.495\linewidth}
  \centering
  \includegraphics[width=\linewidth]{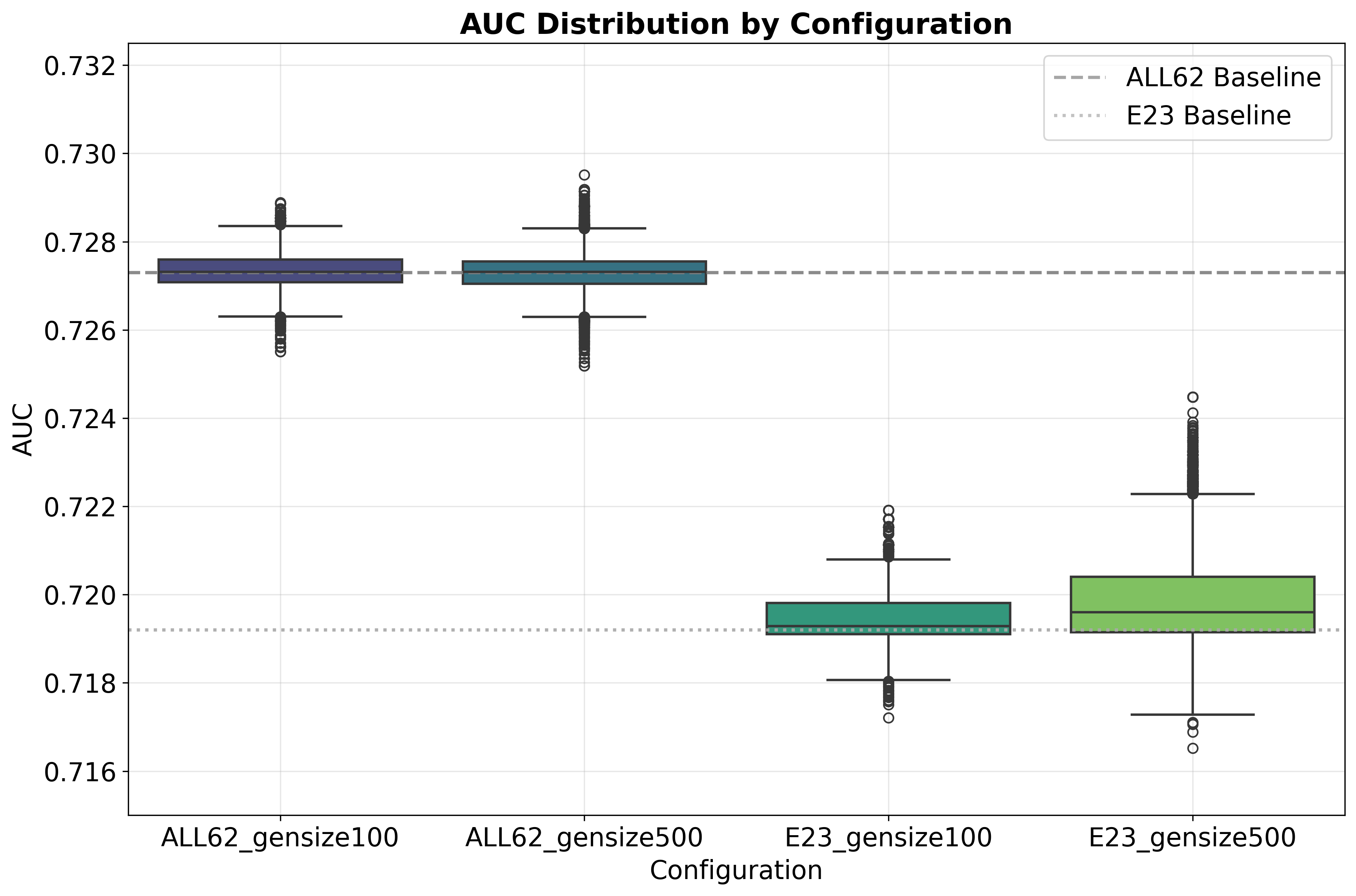}
  \caption{MTG-Jamendo AUC Distribution of all augmented sets\vspace{-3mm}}
  \label{fig:mtg_distribution}
\end{minipage}
\hfill
\begin{minipage}[b]{0.495\linewidth}
  \centering
  \includegraphics[width=\linewidth]{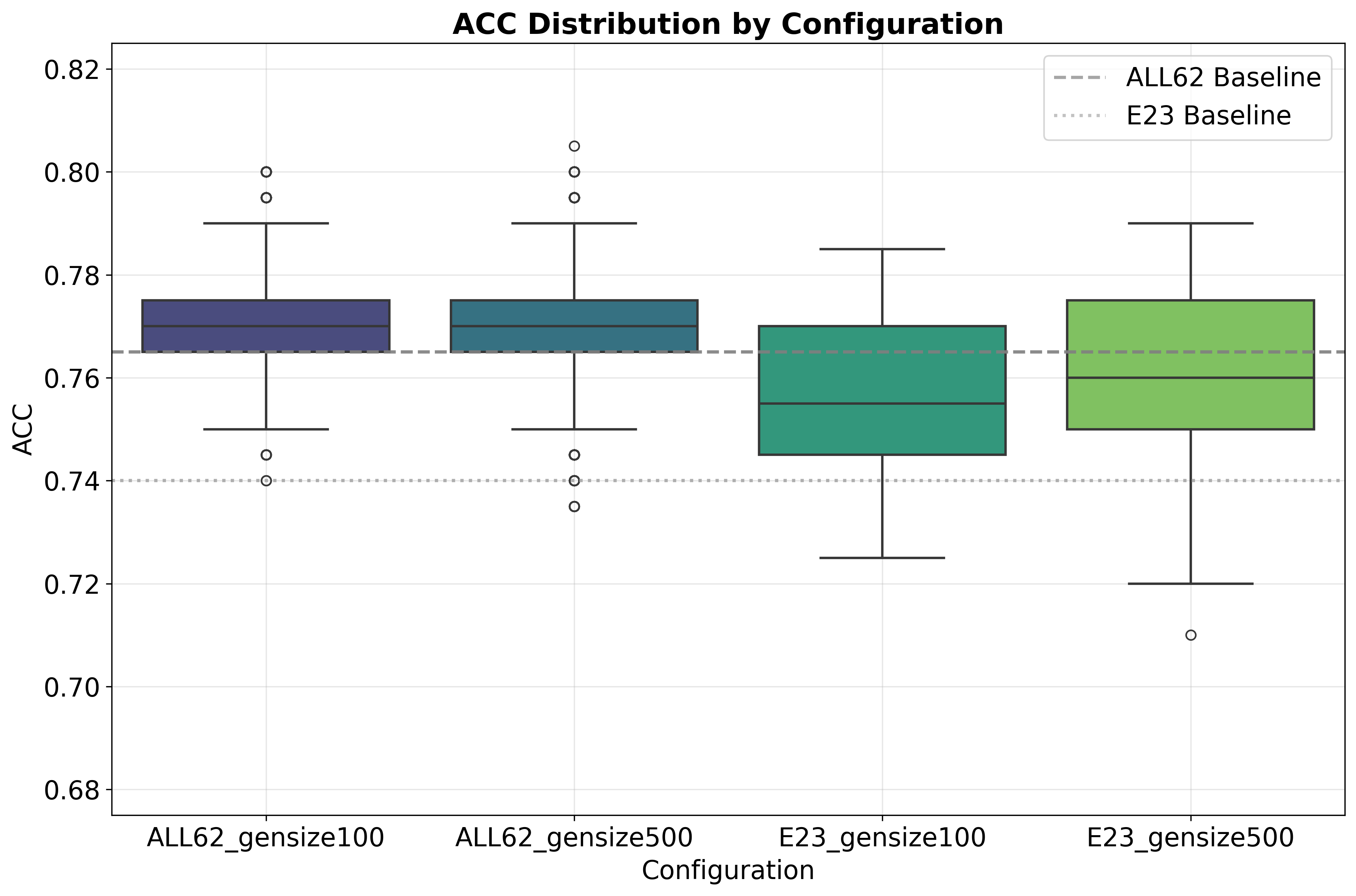}
  \caption{GTZAN Accuracy Distribution of all augmented sets\vspace{-3mm}}  
  \label{fig:gtzan_distribution}
\end{minipage}
\end{figure}

\vspace{-3mm}\subsection{Dataset and Setup\vspace{-2mm}}
Our GP implementation uses populations of 100 and 500 individuals via DEAP~\cite{fortin2012deap}, with termination after 50 generations or early stopping (stagnation: if no improvement for 15 generations; convergence: if fitness variance $< 0.0001$ over 5 generations).
We evaluated the effectiveness of our proposed pipeline against established approaches, which employ XGBoost~\cite{lyberatos2024perceptual,lyberatos2025challenges} without GP enhancement: music mood/theme-tagging on the MTG-Jamendo dataset~\cite{bogdanov2019mtg} and genre-tagging on the GTZAN dataset~\cite{sturm2013gtzan}. The mood/theme subset of the MTG-Jamendo dataset contains 18486 songs with 56 tags. The GTZAN dataset is a collection of 1,000 audio tracks, each 30 sec long, with 100 tracks for each of its 10 featured genres.

Baseline XGBoost models are trained using E23 and ALL62 feature sets, respectively, for both MTG-Jamendo mood/theme tagging and GTZAN genre tagging, following the same train/val/test splits and other setups of the state-of-the-art interpretable methods on Github ~\cite{lyberatos2024perceptual,lyberatos2025challenges}, such as identical XGBoost hyperparameters: 70 estimators with max depth 3 and learning rate 0.1 for MTG-Jamendo multiple binary classification, and max depth 2 with learning rate 0.3 for GTZAN multiclass classification.


\vspace{-4mm}\subsection{GP-Augmented Performance Comparison\vspace{-2mm}}

Our reproduced baselines achieve 0.727 ROC-AUC (MTG-Jamendo) and 0.765 accuracy (GTZAN) using ALL62 features, closely matching but slightly below the original reported performance~\cite{lyberatos2024perceptual,lyberatos2025challenges} of 0.729 ROC-AUC and 0.79 accuracy, respectively. Table~\ref{tab:performance_comparison} presents the best-performing GP-augmented feature sets for deployment on the never-seen test set with stratified bootstrap confidence intervals ($B=2000$, 95\% CI). These results demonstrate two key findings: first, GP consistently improves upon both base feature sets across datasets, with particularly substantial gains on GTZAN (4.0-5.0\% accuracy improvement). This indicates that GP effectively discovers beneficial combinations, regardless of the level of information of the base feature set. Second, our method surpasses state-of-the-art interpretable approaches~\cite{lyberatos2024perceptual,lyberatos2025challenges}) while approaching state-of-the-art deep learning performance (0.781 ROC-AUC on MTG-Jamendo~\cite{kang2025towards}, 0.84 accuracy on GTZAN~\cite{10097236}) with only 5 GP iterations.


To validate that the performance improvements of the best-performing GP-augmented feature sets (Table~\ref{tab:performance_comparison}) reflect systematic effectiveness rather than outlier behavior, Figures~\ref{fig:mtg_distribution} and~\ref{fig:gtzan_distribution} present the complete performance distributions across all GP-augmented feature sets. The key finding is that median GP performance consistently exceeds baselines across all configurations (population size 100 or 500), with narrow inter-quartile ranges indicating stable improvement. While the median gains are small, the findings indicate that most GP-evolved features are beneficial. Deployment models (i.e., those shown in Table 1) occupy the upper tail of an overall improved distribution, rather than being rare outliers. 



\vspace{-4mm}\subsection{GP Improvement Trajectory\vspace{-2mm}}

\begin{figure}[t]
\centering
\begin{minipage}[b]{\linewidth}
\centering
\includegraphics[width=\linewidth]{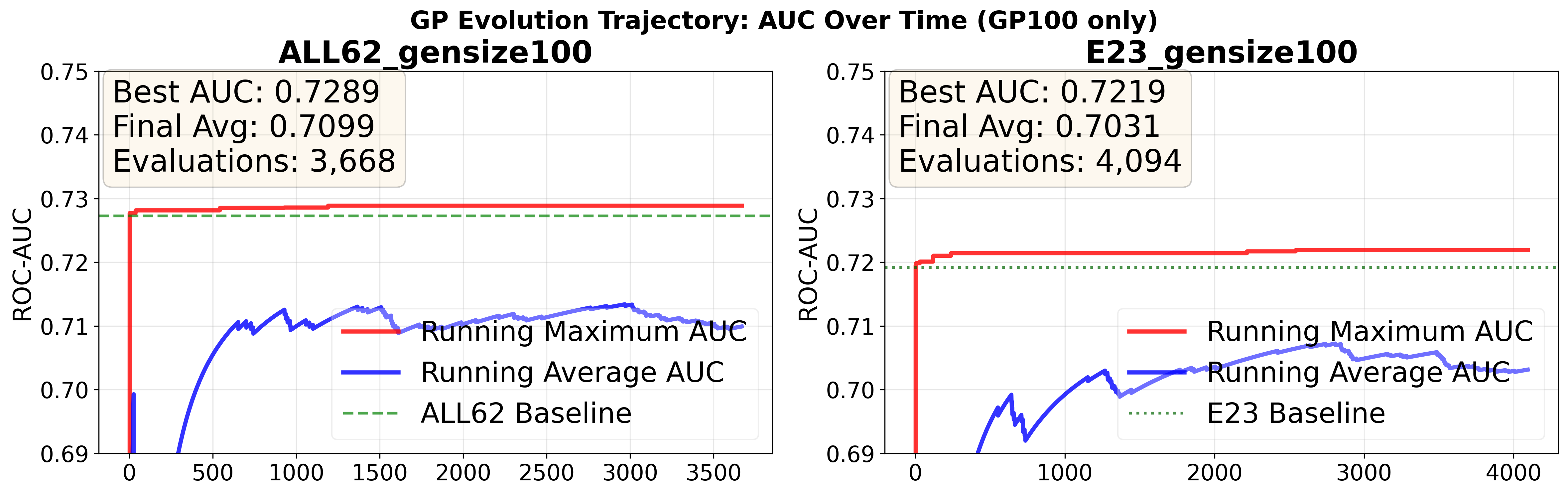}
\caption{\vspace{-2mm}MTG-Jamendo Improvement Trajectory\vspace{-3mm}}
\label{fig:mtg_trajectory}
\end{minipage}

\begin{minipage}[b]{\linewidth}
\centering
\includegraphics[width=\linewidth]{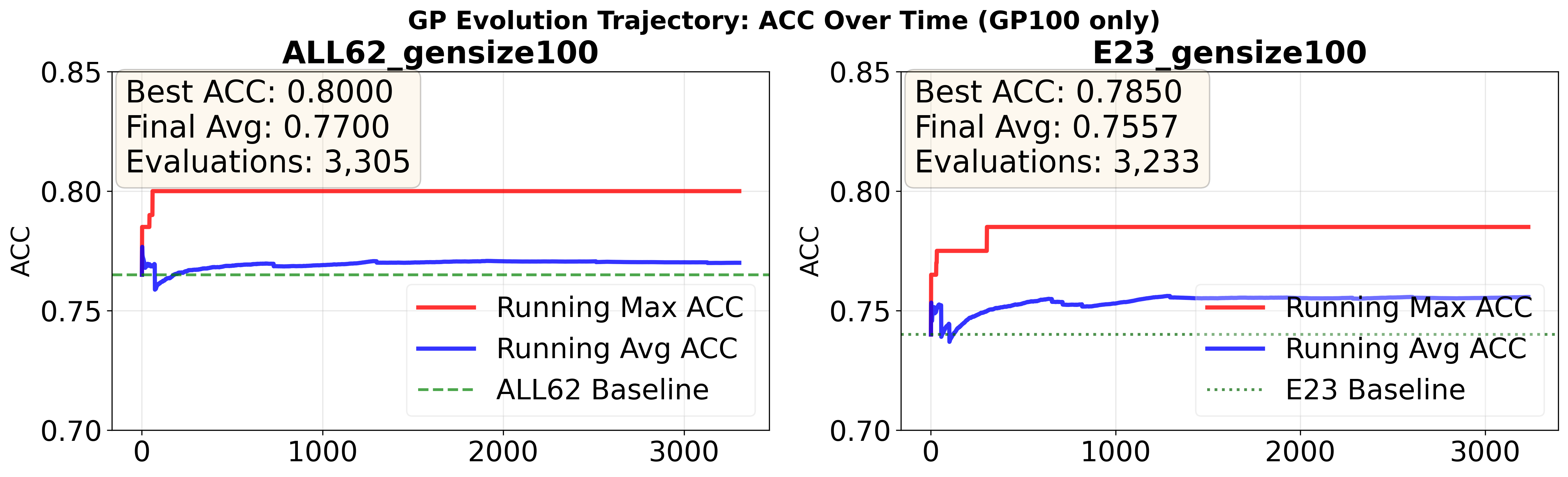}
\caption{\vspace{-2mm}GTZAN Improvement Trajectory\vspace{-3mm}}
\label{fig:gtzan_trajectory}
\end{minipage}
\end{figure}

To assess computational cost-efficiency, Figures~\ref{fig:mtg_trajectory} and~\ref{fig:gtzan_trajectory} present the anytime trajectories of GP feature construction, showing best-so-far and running-average scores versus evaluation count for GP configuration of 100 individuals in population. Across datasets, the best-so-far curves rise steeply during the first few hundred evaluations and then flatten, while the running averages increase more gradually toward a plateau. This improvement pattern remains consistent across population sizes, with GP500 showing similar trajectories (additional plots available in our GitHub repository in Section~\ref{sec:intro}). This characteristic steep-then-plateau pattern reflects evolutionary search dynamics in large combinatorial spaces: early generations explore diverse feature combinations yielding rapid gains, while later refinement produces diminishing returns as the search approaches local optima. Dataset complexity determines convergence rates. Specifically, in GTZAN's 10-class genre classification, our method achieves near-optimal performance within 300 evaluations, while in MTG-Jamendo's challenging 56-tag multilabel prediction, it requires ~1000 evaluations for substantial gains, with the largest improvement (E23-GP500) extending to 8000 evaluations. These improvements are both computationally efficient (5.5 sec per evaluation for MTG-Jamendo, 1.2 sec for GTZAN on RTX 3080Ti) and practically significant, surpassing established interpretable methods while maintaining modest search budgets.

\vspace{-3mm}\section{Interpretability Analysis on GP Features}
\label{sec:expana}

\vspace{-3mm}

The interpretability advantage of our GP approach is demonstrated by analysing the symbolic mathematical expressions of evolved composite features and their feature-feature and feature-operator co-occurrence patterns, extending beyond traditional importance-based methods~\cite{lyberatos2024perceptual, mishra2017local, chen2016xgboost} to reveal how base features should be combined and transformed for optimal tagging performance. Note, base feature names appear in typewriter font (e.g., \texttt{Spectral-Spread}).

\vspace{-4mm}\subsection{Best Features Expressions\vspace{-2mm}}

We analyze composite features from the top-performing evolved expressions to understand base feature interaction mechanisms. Table~\ref{tab:combined_best_features_compact} shows that GP discovers heterogeneous solutions ranging from linear and nonlinear combinations to complex conditional forms, revealing multiple effective pathways for feature combination rather than convergence on a single type. Several low-complexity expressions achieve top performance alongside deeper constructs, consistent with parsimony pressure~\cite{poli2008parsimony} that compact expressions curb bloat and are easier to interpret. For example, the MTG-Jamendo expression $\texttt{Loudness} - \texttt{BPM}^{\texttt{dom\_dom}}$ potentially captures perceived energy through dynamics-tempo interaction modulated by harmonic tension. Moreover, conditional operators could suggest piecewise or categorical feature-label correlations, such as GTZAN's conditional expression using $\texttt{Mode}$ and $\texttt{Chords\_Changes\_Rate}$ could exploit genres' characteristic differences in harmonic rhythm and major/minor tendencies.

\begin{table}[t]
\setlength{\tabcolsep}{0pt} 
\renewcommand{\arraystretch}{0.95}

\begin{tabularx}{\columnwidth}{@{}X@{}}

\caption{Top GP Individual Expressions}
\label{tab:combined_best_features_compact} \\

\hline
\textbf{Dataset: MTG-Jamendo (Base: ALL62, Metric: AUC)} \\
\textbf{AUC 0.730:} \hfill$\max(0, \cos(\text{glob\_glob\_sub}))
 $\\ \hfill $- \cos(\max(0, \text{Rhythm\_Stability}))+ \text{Spectral\_Energyband\_Low}$

\textbf{AUC 0.729:} \hfill \(\text{Loudness} - \text{BPM}^{\text{dom\_dom}}\) \\
\hline

\textbf{Dataset: GTZAN (Base: ALL62, Metric: ACC)} \\
\textbf{ACC 0.805:} \hfill \(\frac{\max(\text{sub\_sub\_dom}, \text{glob\_dom\_dom})}{\text{Danceability}} \;|\;\) \\
\hfill \((\text{glob\_dom}^{\text{Spectral\_Energyband\_Middle\_Low}})^2\)

\textbf{ACC 0.800:} \hfill \(\text{if}\Big(\text{if}(\text{Mode}>0, \cosh(\text{Chords\_Changes\_Rate}), \) \\
\hfill \(\frac{\text{Spectral\_Entropy}}{\text{dom\_dom\_tonic}})>0, \quad \frac{1}{1+e^{-\text{glob\_sub}}}, \;
\frac{1}{\text{Chords\_Number\_Rate}-\text{glob\_dom\_tonic}}\Big)\) \\
\hline

\textbf{Dataset: MTG-Jamendo (Base: E23, Metric: AUC)} \\
\textbf{AUC 0.724:} \hfill \(2 \cdot \text{Length} - 3 \cdot \text{Onset\_Rate} \) \\
\hfill \(- \min(\text{Danceability}, \text{Spectral\_Decrease})\)

\textbf{AUC 0.724:} \hfill \(\tanh\Big(\text{Length} - 2 \cdot \text{Onset\_Rate} - \text{Zerocrossingrate}\Big)\) \\
\hline

\textbf{Dataset: GTZAN (Base: E23, Metric: ACC)} \\
\textbf{ACC 0.790:} \hfill \(\tan(\max(\text{Spectral\_Flux}, \text{Spectral\_Rolloff})) \;|\;\) \\
\hfill \(\frac{1}{1+e^{-(\cosh(\text{Spectral\_Energyband\_Middle\_Low}))^2}}\) \\

\textbf{ACC 0.785:} \hfill \(\text{Chords\_Number\_Rate} \cdot \text{Danceability} \;|\; \) \\
\hfill \(\frac{\text{Onset\_Rate}}{\text{Dynamic\_Complexity}} - \text{Danceability} + \text{Onset\_Rate}\) \\
\hline

\end{tabularx}
\vspace{-4mm}
\end{table}

\vspace{-4mm}\subsection{Base Features and Operation Co-occurrence\vspace{-2mm}}

\begin{figure}[t]
\centering
\begin{minipage}[t]{\linewidth}
  \centering
  \includegraphics[width=\linewidth]{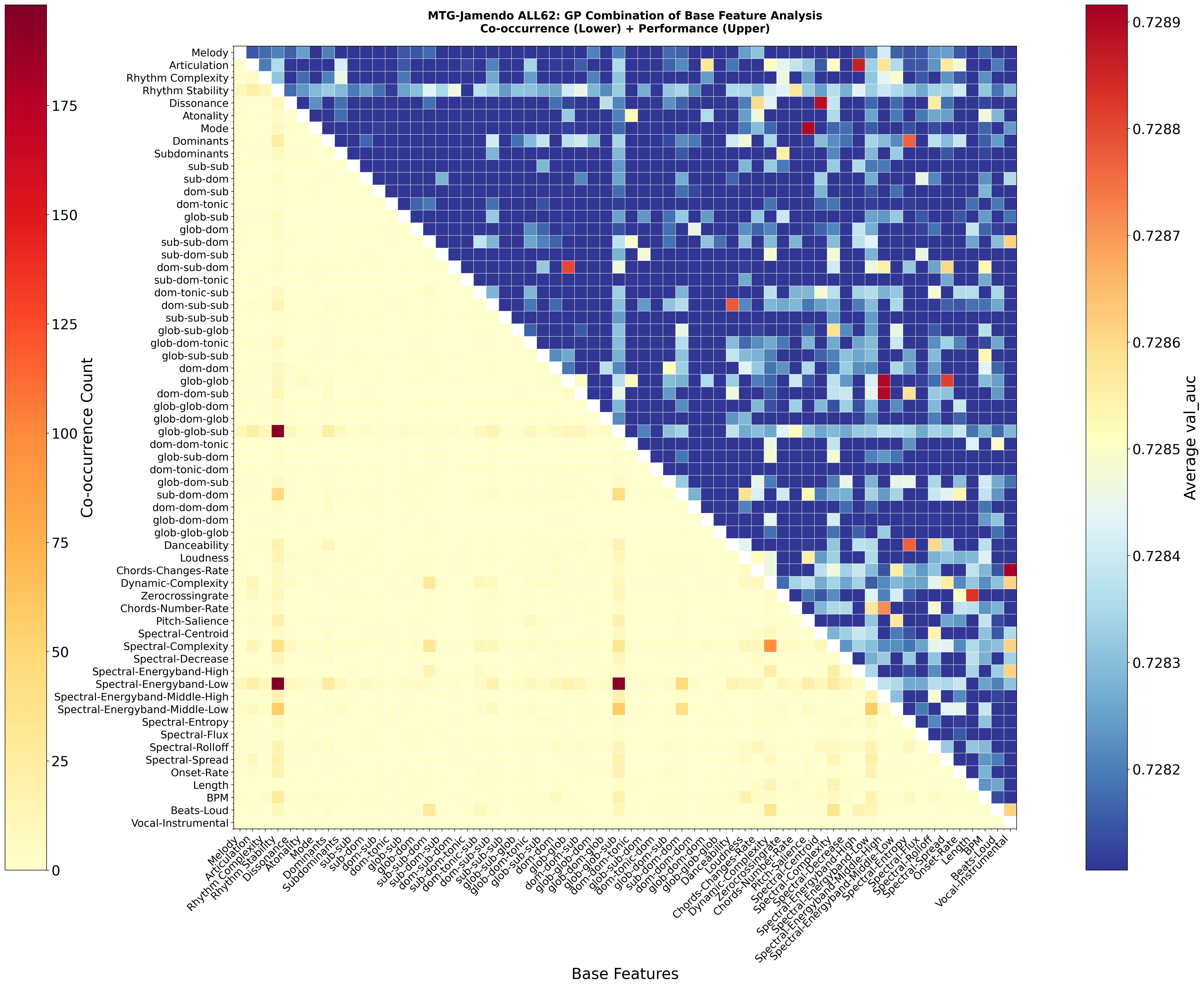}
  \caption{\vspace{-4mm}MTG-Jamendo ALL62 Co-occurrence Matrix}
  \label{fig:mtg_all62_cooccur}
\end{minipage}


\begin{minipage}[t]{\linewidth}
  \centering
  \includegraphics[width=\linewidth]{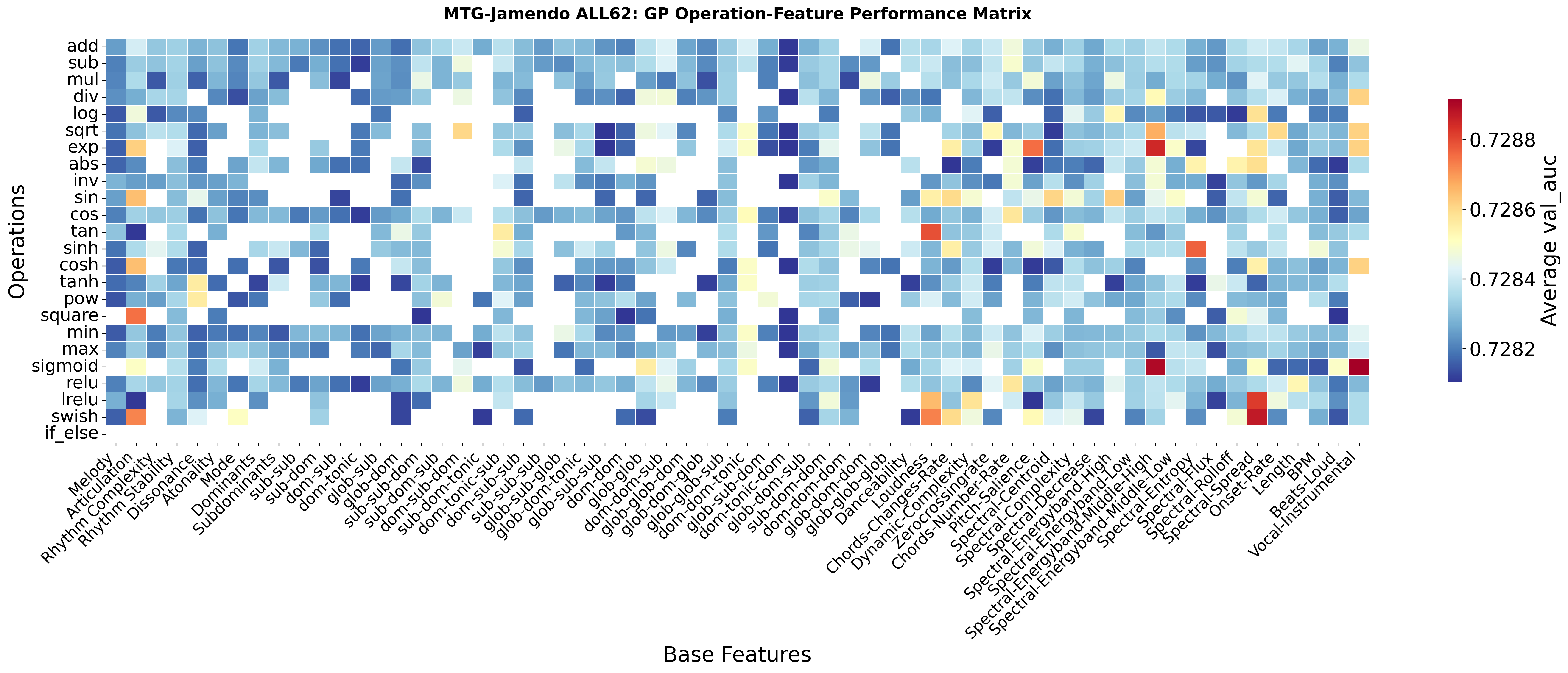}
  \caption{MTG-Jamendo ALL62 Operator-Feature Performance\vspace{-5mm}}
  \label{fig:mtg_all62_opfeat}
\vspace{-4mm}
\end{minipage}
\end{figure}

Co-occurrence analysis reveals task-specific feature synergies with music-theoretic implications. Figure~\ref{fig:mtg_all62_cooccur} shows pairwise co-occurrence (lower triangles) and mean performance conditioned on pair presence (upper triangles) for top-500 evolved expressions. For MTG-Jamendo, \texttt{Spectral-Spread} paired with timbral features appears at moderate frequency with high ROC-AUC, potentially capturing how spectral distribution affects perceived brightness and warmth. Additionally, \texttt{Spectral-\hspace{0pt}Decrease} with \texttt{Beats-Loud} shows mid-range frequency but strong performance, suggesting rhythmic and spectral characteristics jointly contribute to mood discrimination. GTZAN exhibits different synergies. It benefits from harmonic function pairs (e.g., \texttt{dom-\hspace{0pt}tonic-\hspace{0pt}dom} with \texttt{glob-dom-glob} showing high mean accuracy), reflecting genre-distinctive chord progression patterns that GP discovers as harmonically-informed combinations. The frequent co-occurrence of \texttt{Spectral-Entropy} with \texttt{Spectral-Flux} indicates that spectral irregularity measures together enhance genre classification.

Operator-feature patterns shown in Figure~\ref{fig:mtg_all62_opfeat} reveal perceptually-motivated transformations across tasks. For MTG-Jamendo, logarithmic operations on temporal features (\texttt{Danceability}, \texttt{Onset-\hspace{0pt}Rate}, \texttt{Length}) appear frequently with high ROC-AUC, potentially reflecting rhythmic perception's nonlinear nature, consistent with logarithmic scaling observed in many perceptual domains. Sigmoid/swish operations with \texttt{Vocal-\hspace{0pt}Instrumental} features at lower frequency but high performance suggest nonlinear vocal-instrumental balance contributes to mood perception through threshold-like mechanisms. GTZAN shows different patterns. Sigmoid transforms on spectral features (\texttt{Spectral-Flux}, \texttt{Spectral-\hspace{0pt}Rolloff}) achieve near-best accuracy, possibly modeling threshold effects in timbre perception, while power operations with harmony features (\texttt{glob-dom}, \texttt{glob-sub-dom}) further support the genre-harmonic relationship. (Corresponding GTZAN plots are available in our GitHub repository.)



\vspace{-4mm}\section{CONCLUSION}
\label{sec:conclu}

\vspace{-3mm}

This paper proposes a GP pipeline to construct interpretable composite features that augment base feature sets for music tagging. Our method mathematically combines base music features at scale to capture synergistic interactions, thereby achieving representational benefits similar to deep learning–style feature fusion. However, unlike the latter methods, our GP pipeline preserves interpretability. Experiments on MTG-Jamendo and GTZAN datasets show consistent performance gains across base features of varying abstraction levels, with improvements emerging within the first few hundred GP evaluations, and hence with a relatively modest search budget. The evolved best-performing expressions range from linear and non-linear to conditional forms, including low-complexity solutions at top performance, showing the GP’s parsimony pressure produces simpler, more interpretable expressions without sacrificing effectiveness. Analysis of feature-feature and feature-operator co-occurrence patterns reveals that interpretable feature construction extends beyond traditional importance-based methods. 



\vfill\pagebreak


\bibliographystyle{IEEEbib}
\bibliography{refs}

\end{document}